\documentclass[3p,times,twocolumn]{elsarticle}

\usepackage{ecrc,amssymb,amsmath}

\volume{00}

\firstpage{1}

\journalname{Nuclear Physics B Proceedings Supplement}

\runauth{}

\jid{nuphbp}

\jnltitlelogo{Nuclear Physics B Proceedings Supplement}

 \usepackage{amsthm}
 
 \usepackage{subfigure}
 \usepackage{float}
 \usepackage{caption}
 \usepackage{multicol}
 \usepackage{graphicx}
  \usepackage{boxhandler}
  \usepackage[singlelinecheck=off]{caption}
 \newcommand{\bea}{\begin{equation}}
\newcommand{\eea}{\end{equation}}
\newcommand{\be}{\begin{eqnarray}}
\newcommand{\ee}{\end{eqnarray}}
\newcommand{\nn}{\nonumber}
\def\lsim{\mathrel{\mathpalette\@versim<}}
\def\gsim{\mathrel{\mathpalette\@versim>}}

\usepackage[figuresright]{rotating}

\begin{document}

\begin{frontmatter}


\title{Higgs(es) in triplet extended supersymmetric standard model at the LHC}

\dochead{HIP-2014-21/TH}


 \author[label2]{Priyotosh Bandyopadhyay\corref{cor1}}
 \ead{priyotosh.bandyopadhyay@helsinki.fi}

 \author[label2]{Katri Huitu}
 \ead{katri.huitu@helsinki.fi}

 \author[label2]{Asl{\i} Sabanc{\i} Ke\c{c}eli}
 \ead{asli.sabanci@helsinki.fi}

 \address[label2]{Department of Physics, University of Helsinki and Helsinki Institute of Physics,
P.O.Box 64 (Gustaf H\"allstr\"omin katu 2), FIN-00014, Finland}

\cortext[cor1]{Corresponding author}

\begin{abstract}
The recent discovery of  the $\sim 125$ GeV Higgs boson by Atlas and CMS experiments has set strong constraints on parameter space of the minimal supersymmetric model (MSSM). However these constraints can be weakened by enlarging the Higgs sector by adding a triplet chiral superfield. In particular, we focus on the $Y=0$ triplet extension of MSSM, known as TESSM, where the electroweak contributions to the lightest Higgs mass are also important and comparable with the strong contributions. We discuss this in the context of the observed Higgs like particle around 125 GeV and also look into the status of other Higgs bosons in the model. We calculate the Br($B_s \to X_s \gamma$) in this model where three physical charged Higgs bosons and three charginos contribute. We show that the doublet-triplet mixing in charged Higgses plays an important role in constraining the parameter space. In this context we also discuss the phenomenology of light charged Higgs probing $H^\pm_1-W^\mp-Z$ coupling at the LHC.
\end{abstract}


\end{frontmatter}


\section{Introduction}\label{intro}
\vspace*{-0.2cm}
The discovery of the Higgs boson \cite{Higgsd1,Higgsd2} has given us new window in understanding the electroweak symmetry breaking (EWSB) and its underlying theory. The experimental results for Higgs production and decay channels are in very good agreement with the Standard Model (SM) predictions \cite{ZZ1,ATLAS:2013nma} but there are still room for other models. Such models often are motivated by the problems of the SM such as naturalness, the lack of neutrino masses and a dark matter candidate. Supersymmetry (SUSY) removes the famous hierarchy problem and also gives dark matter candidates. 

In the minimal supersymmetric extension of the SM (MSSM), the lightest neutral Higgs mass
is $m_h\leq m_Z$ at the tree-level and the measured Higgs mass can only be achieved with the help of large  radiative corrections. The observation of the $\sim 125$ GeV Higgs thus either leads to a large mixing between the third generation squarks and/or soft masses greater than a few TeV \cite{mssmsd, mssmft}. This pushes the SUSY mass limit to $\gtrsim $ a few TeV for the most constrained scenarios \cite{cmssmstd,cmssmfit}.

Here we consider the triplet supersymmetric extension of Standard Model (TESSM).  This extension helps to accommodate a light Higgs boson around 125 GeV without 
pushing the SUSY mass scale very high. This happens for two reasons, firstly due to the extra tree level contribution from the triplet and secondly it also contributes substantially at 1-loop level. TESSM has an extended Higgs sector which constitutes more than one neutral as well as charged Higgs bosons. In this contribution we will report our analysis of the case where the lightest CP-even neutral scalar is the candidate discovered Higgs boson around $\sim 125$ GeV.  

In section~\ref{mdl} we discuss the model briefly and in section~\ref{status} we present the status of
the $\sim 125$ GeVHiggs  in this model after the Higgs discovery. In section~\ref{pheno} we discuss the charged Higgs phenomenology at the LHC and conclude in section~\ref{concl}.
\vspace*{-0.3cm}
\section{The model}\label{mdl}
\vspace*{-0.2cm}
In TESSM, the field content of the MSSM is enlarged by introducing an SU(2) complex Higgs triplet with zero hypercharge which can be represented as a 2x2 matrix 

 \begin{equation}
 {\bf \Sigma} = \begin{pmatrix}
       \sqrt{\frac{1}{2}}\xi^0 & \xi_2^+ \cr
      \xi_1^- & -\sqrt{\frac{1}{2}}\xi^0
       \end{pmatrix} .
 \end{equation}
Here $\xi^0$ is a complex neutral field, while  $\xi_1^-$ and $\xi_2^+$ are the charged Higgs fields. 
Note that $(\xi_1^-)^*\neq -\xi_2^+$. The triplet part of superpotential  of  the Higgs sector in addition to MSSM is given by 
\be
W_{TESSM}&=&\lambda H_d \cdot \Sigma H_u\, +\, \mu_T Tr(\Sigma^2)
\label{superpot}
 \ee
where $\lambda$ is the coupling of Higgs doublets and triplet and $\mu_T$ is the mass parameter of the triplet.

\vspace*{-0.2cm}
\section{Status of $\sim 125$ GeV Higgs and Br($B_s \to X_s \gamma$)}\label{status}
\vspace*{-0.2cm}
After the electroweak symmetry breaking, TESSM possesses three neutral CP-even Higgses ($h_1$, $h_2$, $h_3$), two CP-odd Higgses ($A_1$, $A_2$) and three charged Higgs bosons ($h^\pm_{1,2,3}$ ). This makes the Higgs phenomenology very rich. The lightest neutral Higgs mass gets an additional contribution from the $\lambda$ term in Eq.~\ref{superpot} at the tree-level. This extra contribution increases for low $\tan{\beta}$ and high $\lambda$ \cite{stefano}
\be\label{mh1}
m^2_{h^0_1}\leq m_Z^2 \left( \cos{2\beta} + \frac{2\lambda^2}{g_1^2+g_2^2} \sin{2\beta} \right)\ ,\quad \tan\beta=\frac{v_u}{v_d}\ ,
\ee
 where  $g_1$ and $g_2$ are the electroweak gauge couplings and $v_u, v_d$ are doublet VEVs.
Due to this additional tree-level contribution, the required loop contributions to have the lightest neutral Higgs mass around $\sim 125$ GeV are decreased so that the required soft SUSY mass scale does not have to be as high as in the case of MSSM.
 
\begin{multicols}{2}
\begin{figure*}
\captionStyle{0}{}
\captionbox{$m_{h_1}$ vs $m_{\tilde{t}_1}$ for the minimal mixing scenarios, where the black and red points represent $\lambda=0.1$; and the green and blue points stand for $\lambda=0.9$ for $\tan\beta=5$.\label{xtb}\cite{trp1}}{\includegraphics[width=3in,height=2.in]{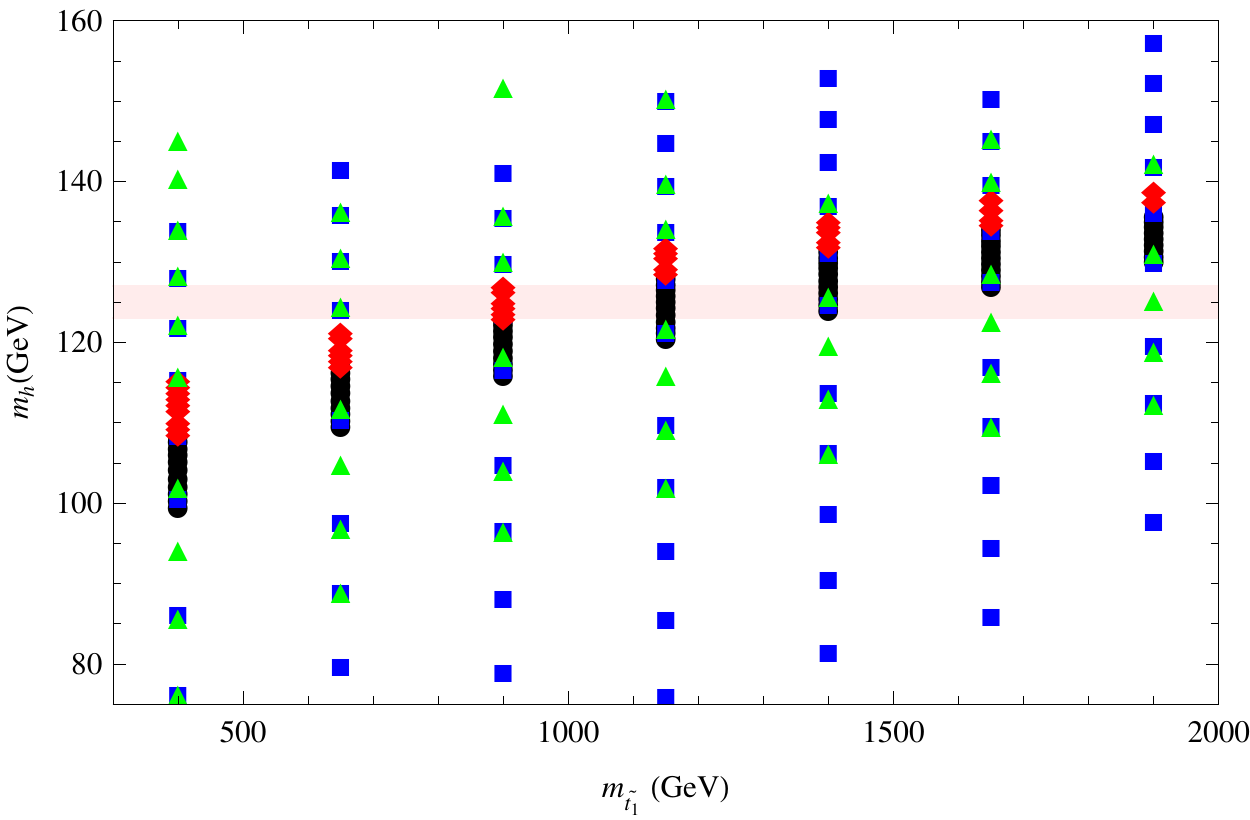}}
\end{figure*}
\end{multicols}
\vspace{-0.5cm}
We calculated the one-loop neutral Higgs masses via Coleman-Weinberg effective potential approach \cite{coleman}.  We found that the contribution of triplet superfield to one-loop Higgs mass is not negligible specially for large $\lambda$ and further reduces the required SUSY mass scale \cite{trp1}.
From Fig.~\ref{xtb} we can see that even for minimal mixing scenario (no-mixing in the stop sector of MSSM), the required SUSY mass scale is less than TeV.  Thus $\sim 125$ GeV Higgs in a supersymmetric scenario does not imply that supersymmetry will be discovered beyond few TeV and we have to look for the direct SUSY searches. 

The important feature of the triplet is that it does not couple to the SM fermions directly (see Eq.~\ref{superpot}.). Thus it diminishes the strength of the SM fermion-Higgs coupling through the doublet-triplet mixing governed by the $\lambda$ term in the superpotential.  The mixing certainly changes the decay widths of Higgs to gluon pair as well as Higgs to di-photon \cite{trp2}. In TESSM there are three charged Higgs bosons and three charginos which contribute to the rare decays such as $B_s \to X_s \gamma$. If the corresponding charged Higgses or charginos are triplet type they will not contribute to the decay process. Thus such doublet-triplet mixing is very crucial. Generally the charged Higgs and chargino diagrams come with different sign and depending on their doublet content, they partially cancel. 

\begin{figure*}
\captionStyle{}{1}
\vskip -30 pt
\captionbox{The doublet-triplet in lightest charged Higgs mass eigenstate with the variation of $\mu_D$ for $\lambda=0.9$ and $\tan{\beta}=5$.\label{chhiggsmixing}\cite{trp1}}
{\includegraphics[width=3.2 in,height=2.5 in]{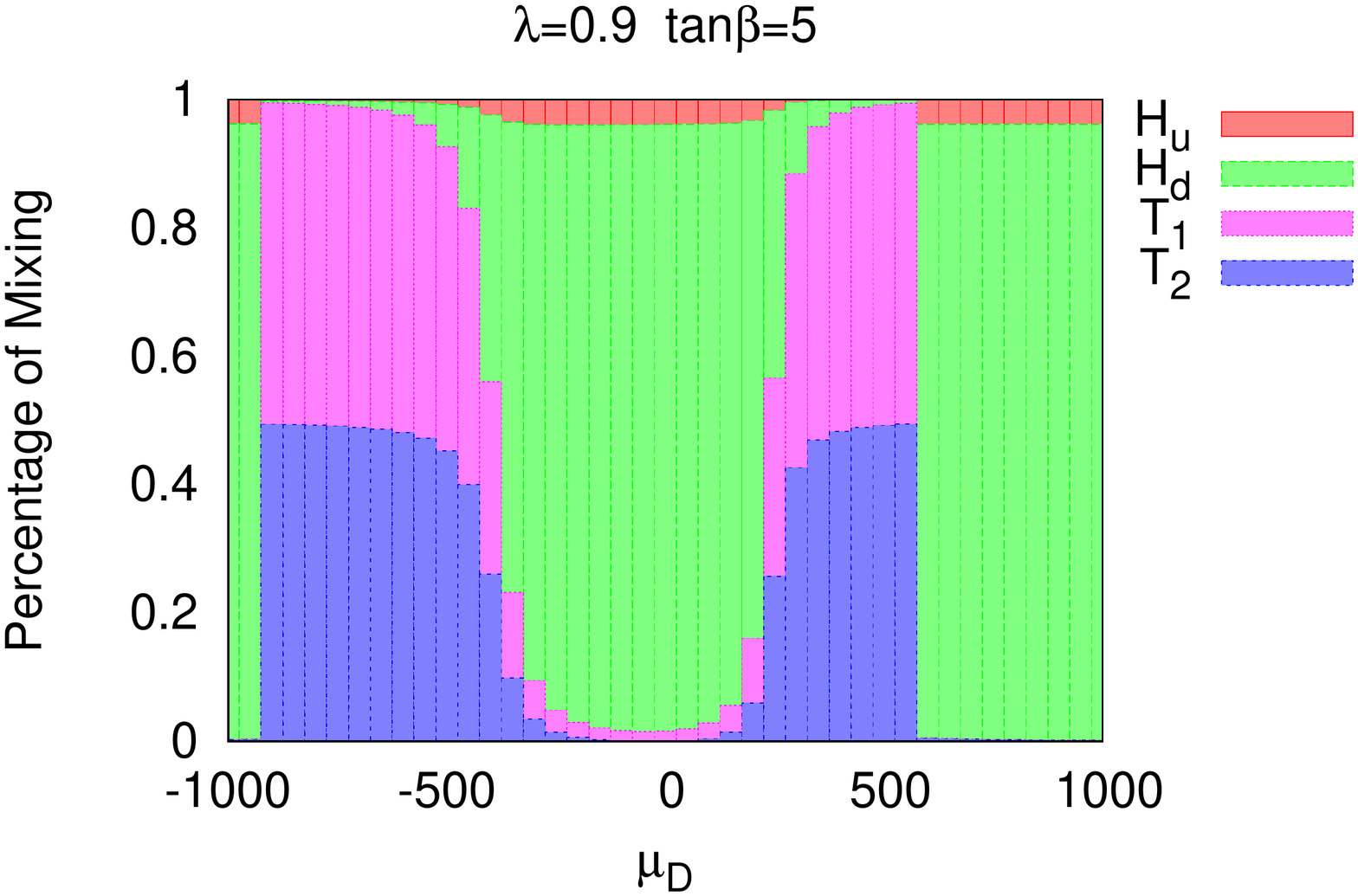}}
\end{figure*}

In Fig.~\ref{chhiggsmixing} we show the doublet and triplet structure of the light charged Higgs as a 
function of $\mu_D$, the mixing parameter between the two Higgs doublets ($H_u$, $H_d$) for $\lambda=0.9$ and $\tan{\beta}=5$. It can be seen that the mixing could be substantial and can lead to cancellation between charged Higgs and chargino diagrams at a different parameter space  than in the MSSM.  These two reasons lead to allowed parameter space quite different than MSSM or 2HDM. We have also calculated Br($B_s\to X_s \gamma$) at NLO in this model and showed that the $2\sigma$ allowed 
region constrains the high $\lambda$ region of parameter space preferred by naturality\cite{trp2}. 
\begin{multicols}{2}
\end{multicols}
\vspace*{-1.2 cm}
\section{Phenomenology at the LHC}\label{pheno}
\vspace*{-0.2cm}
We have seen that TESSM can accommodate a $\sim 125$ GeV Higgs consistent with the Higgs data and the constraints from Br($B_s\to X_s \gamma$) \cite{trp1, trp2}. It would be interesting to see if we can probe the triplet nature of the model by some other properties of it. 

In any hypercharge $Y=0,\pm 2$ triplet extended model, the charged Higgs bosons couple
to $ZW^\pm$ at tree level. Such a coupling is induced only at loop level in MSSM or 2HDM. For $Y=0$ triplet extended model the coupling can be written as
\vspace*{-0.2cm}
\be
g_{h_{i}^\pm W^\mp Z}=-\frac{1}{2} i g_2 \left(g_1 \sin\theta_w (v_u \mathcal{R}_{(i+1)1}-v_d \mathcal{R}_{(i+1)2})\right) \\ \nn
-\frac{1}{2} i g_2 \left(\sqrt{2} g_2 v_T \cos\theta_W (\mathcal{R}_{(i+1)3}+\mathcal{R}_{(i+1)4})\right).
\ee 
where $\theta_W$ is the Weinberg angle and the triplet VEV $v_T\lesssim 5$ GeV by the EW $\rho$ parameter \cite{pdg}. $\mathcal{R}_{ij}$ is the rotation matrix that relates the physical charged Higgs mass eigen state  with the charged Higgs fields as, $h_i= R_{ij} H_j$. The  ${h_{i}^\pm W^\mp Z}$ coupling goes to zero for $v_T=0$ to preserve $U(1)_{em}$ gauge symmetry. A non-zero triplet VEV ($v_T$) leads to a unique coupling that carry the triplet information and can be obtained through charged Higgs phenomenology. 

Recent search for light charged Higgs at the LHC for 2HDM and MSSM looked for $pp\to t\bar{t}$ and assumed Br($H^\pm\to \tau \nu$) to be 100\% \cite{chbnd}.  This lower bound certainly changes with the non-standard decays of the light charged Higgs. In TESSM if the lighter charged Higgs is triplet type then it can decay to $ZW^\pm$ and when kinematically allowed gets strong competition from $tb$ mode \cite{trp3}. 

As the triplet type charged Higgs does not couple to fermions, the dominant production modes  in MSSM i.e., $g g\to h^\pm_1 t b$ and $b g\to h^\pm_1 t$ are no longer dominant for heavy triplet type charged Higgs. For $m_{h^\pm_1}< m_t$, $pp\to t\bar{t}$ is not the primary production process at the LHC in which ATLAS/CMS searched for light charged Higgs \cite{chbnd}. Certainly we need to look for other production mechanisms and due to the triplet nature of the charged Higgs they have to be electroweak. The charged Higgs pair productions $h^\pm_i h^\mp_j$, associated gauge boson productions: $h^\pm_i W^\pm$, $h^\pm_i Z$ and associated neutral Higgs productions i.e. $h^\pm_i \phi$ ($\phi:h_{1,2,3}, \, A_{1,2}$) could be important. In addition we have vector boson fusion channel giving charged Higgs due to no-zero ${h_{i}^\pm W^\mp Z}$ which is absent in MSSM.  

As the charged Higgs can decay to $ZW^\pm$, it can lead to multi-leptonic final states. The charged Higgs productions in pair or in association with neutral Higgs and gauge bosons can lead to multi-lepton $+\tau$ and multi-lepton + jets final states. All these modes are quite interesting in probing the light charged Higgs boson at the LHC 14 TeV which can provide us the earliest hint about TESSM \cite{trp3}. Reconstructing $ZW^\pm$ invariant mass via constructing $\ell\ell jj$ invariant mass can give us the 
charged Higgs mass peak. 
\vspace*{-0.5cm}
\section{Conclusions}\label{concl}
TESSM gives an option where we can accommodate a $\sim 125$ GeV Higgs and still expect that
the SUSY scale is not very high. The charged Higgs phenomenology in TESSM is very interesting and can give us the key features of a triplet. In order to pin down the parameter space we need to use the direct and indirect experimental bounds as well as to look for various possible final states. We hope that LHC at 14 TeV can shed some light on its phenomenological searches.
\vspace*{-0.2cm}






\end{document}